\documentclass[11pt,a4paper]{article}
\usepackage{amsmath}
\usepackage{cite}
\usepackage{graphicx}
\usepackage{amsfonts}
\usepackage{amssymb}
\usepackage{psfig}
\def\bt{\beta}

\def\bt'{\beta'}

\begin{document}
\bigskip
%\today
\hfill\hbox{SISSA 60/2004/FM} \vspace{2cm}

\begin{center}
{\Large \textbf{Form factors in the massless coset models\\
\vspace{0.5cm} \Large \textbf{$su(2)_{k+1}
\otimes su(2)_k /su(2)_{2k+1}$}}\\
\vspace{0.5cm}
\Large{\textbf{Part I}}}\\
\vspace{1.2cm} {\Large Paolo Grinza$^{a,b}$ \footnote{
\textsf{grinza@lpm.univ-montp2.fr}} and B\'en\'edicte
Ponsot$^{a,c}$ \footnote{\textsf
{ponsot@fm.sissa.it, benedicte.ponsot@anu.edu.au}}} \\
\vspace{0.7cm} {\it  $^{a}$International School for Advanced Studies (SISSA),\\
Via Beirut 2-4, 34014 Trieste, Italy\\
INFN sezione di Trieste}\\
\vspace{0.7cm} {\it $^{b}$ Laboratoire de Physique Math\'ematique,
Universit\'e Montpellier II,\\
 Place Eug\`ene Bataillon, 34095
Montpellier Cedex 05, France}\\
\vspace{0.7cm} {\it $^{c}$Department of Theoretical Physics,\\
Research School of Physical Sciences and Engineering,\\
Australian National University,\\
Canberra, ACT 0200, Australia}
 \vspace{1.2cm}
\end{center}

\begin{abstract}
Massless flows between the coset model $su(2)_{k+1} \otimes
su(2)_k /su(2)_{2k+1}$ and the minimal model $M_{k+2}$ are studied
from the viewpoint of form factors. These flows include in
particular the flow between the Tricritical Ising model and the
Ising model. Form factors of the trace operator with an arbitrary
number of particles are constructed, and numerical checks on the
central charge are performed with four particles contribution.
Large discrepancies with respect to the exact results are observed
in most cases.
\end{abstract}
\begin{center}
\end{center}

\newpage
\section*{Introduction}
The general problem of finding a suitable description of
Renormalization Group (RG) flows between different non-trivial
critical point in Quantum Field Theory is an open and appealing
problem. In this respect 2$d$ integrable QFTs provide a privileged
framework where there is the actual possibility of finding a
detailed description of such flows. As a matter of fact, one of
the remarkable consequences of integrability is the knowledge of
the exact $S$-matrix which allows to calculate exact
form-factors\cite{KW,BKW,smirnov2}. From the latter one can
reconstruct correlators by means of the spectral expansion which
provides a non-perturbative representation of them. Such a
representation has often been very accurate in the past~-though it
was noticed some time ago in \cite{CAF} that this common belief,
even for massive theories, is too optimistic in general- and
allows, for example, to recover the conformal data using the
$c$-theorem sum rule \cite{Zamolodchikov:1986gt, Cardy:1988tj}
which gives the difference between the central charges of the UV
and IR fixed points. Integrable RG flows which limit in the
infrared is a non-trivial CFT can be described by massless
excitations, for which an exact scattering matrix can be found.
More precisely, we deal with right and left movers and three
different types of scatterings, associated to right-right,
left-left, and right-left interactions. The right-right and
left-left $S$-matrices, whose definition is formal, are
independent of the RG scale, and are solely characterized by the
properties of the IR fixed point CFT. On the contrary, the
right-left scattering is quite rigorously defined\cite{Z2}. It
becomes trivial in the IR limit, thus the left and right movers
decouple, and we obtain the IR CFT. In \cite{DMS}, using the
$S$-matrix proposed in \cite{Z2}, a few form-factors\footnote{We
refer the reader to \cite{DMS} for a discussion on form-factors in
massless theories.} of some local and non local operators were
constructed for the simplest model describing the flow
\cite{Z4,LC,KMS} between the Tricritical Ising model (TIM) and the
Ising model (IM), where the supersymmetry is spontaneously broken.
Remarkably enough, the numerical results performed in \cite{DMS}
show that the knowledge of the form-factor with the lowest number
of intermediate particles is enough to give quite an accurate
approximation for the correlation function at almost all RG scale.
Of course it is desirable to figure out whether these spectacular
features remain valid in less trivial massless flows. In this
respect, one has to take into account results obtained recently in
\cite{P2} about the calculation of form factors in a {\it massive}
integrable model of QFT called the $SS$ model \cite{Fateev}, as
well as in its RSOS restrictions. The numerical checks performed
in \cite{P2} show in particular the two-particles approximation of
the $c$-theorem sum rule fails to give the usual good
approximation of the exact results, showing instead large
discrepancies (about 20-25\%, see below).

In this article we will consider the construction of form factors
of the trace operator for an arbitrary number of intermediate
particles in the family of massless flows \cite{CSS} between the
UV coset models \cite{GKO} $su(2)_{k+1} \otimes su(2)_k
/su(2)_{2k+1}$ and the minimal models $M_{k+2}$ in the IR. These
flows, whose direction is given by the irrelevant operator
$T\bar{T}$, include in particular for $k=1$ the above mentioned
flow from TIM to IM, and for $k=+\infty$ the massless flow from
the Principal Chiral Model at level 1 ($\mathrm{PCM}_1$) to the
$SU(2)_1$ WZNW model \cite{ZZ}.

We recall that the form factors of the trace operator were
actually first obtained in \cite{MS}  in a different
representation from the one presented in this article; the
representation proposed here is simpler. Let us mention also that
no numerical checks were performed in \cite{MS}.

This article is organized as follows: in section \ref{one}, we recall
known facts on the construction of form factors in the Sine-Gordon
model and its RSOS restriction, that we will need in section \ref{two},
where we construct the form factors of the trace operator in the
massless flows mentioned above. Finally, in sect.~\ref{three}
we perform checks on the four
particle form factor by making a numerical estimation of the
variation of the central charge between the UV and the IR CFTs; we
will see that the accuracy becomes poorer and poorer as one
increases $k$, in an even worst manner than what was first
observed in the massive case \cite{P2}, and we will provide a
reasonable explanation for such a loss of accuracy.

\section{Form factors in the sine-Gordon model and RSOS restriction}
\label{one}
In this section we recapitulate known results on form factors in
the SG model in the repulsive regime.\\
The Sine-Gordon model alias the massive Thirring model is defined
by the Lagrangians:
\begin{eqnarray}
\mathcal{L}^{SG}&=&\frac{1}{2}(\partial_{\mu}\varphi)^2+\frac{\alpha}{\beta^2}(\cos \beta\varphi-1),\nonumber\\
\mathcal{L}^{MTM}&=&\bar{\psi}(\mathrm{i}\gamma\partial-M)\psi-\frac{1}{2}g(\bar{\psi}\gamma_{\mu}\psi)^2,
\end{eqnarray}
respectively. The fermi field $\psi$ correspond to the soliton and
antisoliton and the bose field $\varphi$ to the lowest breather
which is the lowest soliton antisoliton bound state.
 The relation between the coupling constants was found in \cite{Coleman}
 within the framework of perturbation theory:
$$
p\equiv\frac{\beta^2}{8\pi-\beta^2}=\frac{\pi}{\pi+2g}.
$$
The two particles $S$-matrix of the SG model \cite{ZZ2} contains
the following scattering amplitudes: the two-soliton amplitude
$a_p(\theta)$, the forward and backward soliton anti-soliton
amplitudes $b_p(\theta)$ and $c_p(\theta)$:
\begin{eqnarray}
b_p(\theta)=
\frac{\sinh\theta/p}{\sinh(\mathrm{i}\pi-\theta)/p}a_p(\theta),
\quad c_p(\theta)=\frac{\sinh \mathrm{i}\pi/p } {\sinh
(\mathrm{i}\pi-\theta)/p }a_p(\theta),\nonumber
\end{eqnarray}
\begin{eqnarray}
a_p(\theta)=\mathrm{exp}\;
\int_{0}^{\infty}\frac{dt}{t}\frac{\sinh\frac{1}{2}(1-p) t \sinh
\frac{t\theta}{\mathrm{i}\pi}}{\cosh \frac{t}{2}\sinh\frac{1}{2}p
t}.\nonumber
\end{eqnarray}
This $S$-matrix satisfies the Yang-Baxter equation as well as the
unitarity condition:
$$
S_p^{SG}(\theta)S_p^{SG}(-\theta)=1,
$$
which can be rewritten for the amplitudes as:
$$
a_p(\theta)a_p(-\theta)=1,\quad
b_p(\theta)b_p(-\theta)+c_p(\theta)c_p(-\theta)=1.
$$
The crossing symmetry condition reads for the amplitudes:
$$
a_p(\mathrm{i}\pi-\theta)=b_p(\theta),\quad
c_p(\mathrm{i}\pi-\theta)=c_p(\theta).
$$
The repulsive regime (with no bound states) corresponds to the condition $p>1$.\\
The form factors $f(\theta_1,\cdots,\theta_{2n})$ of a local
operator\footnote{Here we shall exemplify form factors with an
even number of particles
 only. For form factors with an odd number of particles, we refer the reader to \cite{BFKZ}.} in the SG
model are covector valued functions that satisfy a system of
equations \cite{smirnov2}, which consist of a Riemann-Hilbert
problem:
\begin{eqnarray}
&&f(\theta_1,\cdots,\theta_{i},
\theta_{i+1},\cdots,\theta_{2n})S_p^{SG}(\theta_i-\theta_{i+1})=
f(\theta_1,\cdots,\theta_{i+1},
\theta_i,\cdots,\theta_{2n}),\nonumber \\
&& f(\theta_1,\cdots,\theta_{2n-1},\theta_{2n}+2\mathrm{i}\pi)=
 f(\theta_{2n},\theta_1,\cdots,\theta_{2n-1}),\nonumber
\end{eqnarray}
and a residue equation at $\theta_1=\theta_{2n}+\mathrm{i}\pi$ :
\begin{eqnarray}
\mathrm{res}f(\theta_1,\cdots,\theta_{2n})=-2\mathrm{i}\;
f(\theta_2,\cdots,\theta_{2n-1})
\left(1-\prod_{i=2}^{2n-1}S_p^{SG}(\theta_{i}-\theta_{2n})\right)
e_0,\quad e_0=s_{1}\otimes \bar{s}_{2n} + \bar{s}_1\otimes s_{2n},
\nonumber
\end{eqnarray}
where $s$ (or +) corresponds to the solitonic state (highest
weight state), and $\bar{s}$ (or -) to the antisolitonic state.
The equations written above do not refer to any particular local
operator. It is one of the difficulties of the form factor
approach to identify solutions of these equations with specific
operators.

Form factors containing an arbitrary number of particles for the
energy momentum tensor were first constructed in \cite{smirnov2}
by Smirnov. Below, we make the choice to present the posterior
construction presented in \cite{BFKZ,BK} by Babujian and Karowski
{\it et. al}. We refer the reader
 to these two references for more details.

We first introduce the minimal form factor $f_p(\theta_{12})$ of
the SG model:
 it satisfies the relation
  $$f_p(\theta)=-f_p(-\theta)a_p(\theta)=f_p(2\mathrm{i}\pi-\theta),$$
and reads explicitly
\begin{eqnarray}
f_p(\theta)=-\mathrm{i}\sinh \frac{\theta}{2}
f^{min}_p(\theta)=-\mathrm{i}\sinh \frac{\theta}{2}\exp
\int_{0}^{\infty}\frac{dt}{t}\frac{\sinh\frac{1}{2}(1-p)t}{\sinh\frac{1}{2}p
t\cosh\frac{1}{2}t} \frac{1-\cosh
t(1-\frac{\theta}{\mathrm{i}\pi})}{2\sinh t}. \nonumber
\end{eqnarray}
Its asymptotic behaviour when $\theta \to \pm \infty$ is given by
$ f_p(\theta)\sim \mathcal{C}_p\;
e^{\pm\frac{1}{4}(\frac{1}{p}+1)(\theta-\mathrm{i}\pi)} $, with
the constant
\begin{eqnarray}
\mathcal{C}_p=\frac{1}{2}\exp
\frac{1}{2}\int_{0}^{\infty}\frac{dt}{t}
\left(\frac{\sinh\frac{1}{2}(1-p)t}{\sinh\frac{1}{2}p
t\cosh\frac{1}{2}t\sinh t}-\frac{1-p}{p t}\right).\nonumber
\end{eqnarray}
It is proposed in \cite{BFKZ} that form factors in SG can be
 written\footnote{This representation holds whether operators are
 local or not, topologically neutral or not.}:
\begin{eqnarray}
f(\theta_1,\dots,\theta_{n})=N_{n}
\prod_{i<j}f_p(\theta_{ij})\int_{C_{\theta}}du_1 \dots
\int_{C_{\theta}}du_m\; h_p(\theta,u)
 p_n(\theta,u)\Psi^{p}(\theta,{u}),
\label{ff}
\end{eqnarray}
where we introduced the scalar function (completely determined by
the $S$-matrix)
\begin{eqnarray}
h_p(\theta,u)=\prod_{i=1}^{2n}\prod_{j=1}^{m}\phi_p(\theta_{i}-u_j)
\prod_{1\le r<s\le m}\tau_p(u_r-u_s),\nonumber
\end{eqnarray}
with
$$
\phi_p(u)=\frac{1}{f_p(u)f_p(u+\mathrm{i}\pi)},\quad
\tau_p(u)=\frac{1}{\phi_p(u)\phi_p(-u)}.
$$
$\Psi^p(\theta,{u})$ is the Bethe ansatz state covector: we first
define the monodromy matrix $T_p$ as
\begin{displaymath}
\left(\begin{array}{cc}
A(\theta_1,\dots,\theta_n,u) & B(\theta_1,\dots,\theta_n,u)\\
C(\theta_1,\dots,\theta_n,u) & D(\theta_1,\dots,\theta_n,u)\\
\end{array}\right) \equiv
T_p(\theta_1,\dots,\theta_n,u) = S^{SG}_p(\theta_1-u)\dots
S^{SG}_p(\theta_{n}-u),
\end{displaymath}
the definition of the Bethe ansatz covector is given by
$$
\Psi^p(\theta,{u})= \Omega_{1\dots n}\;
C(\theta_1,\dots,\theta_n,u_1)\dots
C(\theta_1,\dots,\theta_n,u_m),
$$
where $\Omega_{1\dots n}$ is the pseudo vacuum consisting only of
solitons
$$
\Omega_{1\dots n} = s\otimes\dots \otimes s.
$$
The number of integration variables $m$ is related to the
topological charge $q$ ($q\in \mathbb{Z}$) of the operator
considered and the number $n$
 of particles through the relation
\begin{eqnarray}
q=n-2m. \label{charget}
\end{eqnarray}
For example, for $n=2$ and $q=0$,
$$
\Psi^p(\theta_1,\theta_2,u)=\Psi^p_{+-}(\theta_1,\theta_2,u)+\Psi^p_{-+}(\theta_1,\theta_2,u)=
b(\theta_1-u)c(\theta_2-u)s_1\otimes\bar{s}_2+c(\theta_1-u)a(\theta_2-u)\bar{s}_1\otimes
s_2.
$$
It is important to have in mind that the function
$p_{n}(\theta,u)$ is the {\it only} ingredient in formula
(\ref{ff}) which depends on the operator considered. Different
operators will differ by their $p$-function. If the operator is
chargeless, the form factors contain an even number of particles,
and if in addition the operator is local, then the $p$-function
satisfies the conditions\footnote{We consider here only the case
where the operator is of bosonic type and the particles are of
fermionic type. If both are fermionic, there is an extra statistic
factor to be taken into account, see \cite{BK}.}:
\begin{enumerate}
\item
 $p_{2n}(\theta,u)$ is a polynomial in $e^{\pm
u_j}$, $(j=1,\dots ,n)$ and $p_{2n}(\theta,u)=
p_{2n}(\dots,\theta_i-2\mathrm{i}\pi,\dots,u)$
\item
$p_{2n}(\theta_1=\theta_{2n}+\mathrm{i}\pi,\dots\theta_{2n};u_1\dots
u_n=\theta_{2n})= p_{2n-2}(\theta_2\dots\theta_{2n-1};u_1\dots
u_{n-1})+\tilde{p}^{1}(\theta_2\dots\theta_{2n-1}),
$\\
$p_{2n}(\theta_1=\theta_{2n}+\mathrm{i}\pi,\dots\theta_{2n};u_1\dots
u_n=\theta_{2n}\pm \mathrm{i}\pi)=
p_{2n-2}(\theta_2\dots\theta_{2n-1};u_1\dots
u_{n-1})+\tilde{p}^{2}_{\pm}(\theta_2 \dots \theta_{2n-1}), $\\
where $\tilde{p}^{1,2}(\theta_2 \dots \theta_{2n-1})$ are
independent of the integration variables.
\item
$p_{2n}(\theta,u)$ is symmetric with respect to the $\theta$'s and
the $u$'s.
\item
$p_{2n}(\theta+\ln\Lambda,u+\ln\Lambda)=\Lambda^{\mathrm{s}}\;p_{2n}(\theta,u)$
where s is the Lorentz spin of the operator.
\end{enumerate}
Finally, the integration contours $C_{\theta}$ consist of several
pieces for all integration variables $u_j$~: a line from $-\infty$
to $\infty$
 avoiding all poles such that
$\mathrm{Im}\theta_i-\pi-\epsilon<\mathrm{Im}
u_j<\mathrm{Im}\theta_i-\pi,$ and clockwise oriented circles
around the poles (of the $\phi(\theta_i-u_j)$) at $\theta_i=u_j$,
$(j=1,\dots,m)$.
\paragraph{\it Trace of the energy momentum tensor.}
The trace operator is a spinless and chargeless local operator.
 Its $p$-function is \cite{BK}:
\begin{eqnarray}
p_{SG}^{\Theta}(\theta,u)=-\left(\sum_{i=1}^{2n}e^{-\theta_i}\sum_{j=1}^{n}e^{u_j}-
 \sum_{i=1}^{2n}e^{\theta_i}\sum_{j=1}^{n}e^{-u_j}\right).
 \label{trace}
 \end{eqnarray}
The residue equation gives the following relation for the
normalization $N_{2n}^{\Theta}$ (see \cite{BFKZ}):
\begin{eqnarray}
N_{2n}^{\Theta}=N_{2n-2}^{\Theta}\frac{\left(f^{min}_p(0)\right)^2}{4n\pi}\quad
\to N_{2n}^{\Theta}=
N_{2}^{\Theta}\frac{1}{n!}\left(\frac{\left(f^{min}_p(0)\right)^2}{4\pi}\right)^{n-1}.
\label{normalization}
\end{eqnarray}
The two particles form factor can be computed explicitly:
\begin{eqnarray}
f_{SG}^{\Theta}(\theta_{12})= \frac{2\pi
N_2^{\Theta}}{\mathcal{C}_p^4}f_p(\theta_{12})\frac{\cosh
\frac{\theta_{12}}{2}}{\sinh\frac{1}{2p}(\mathrm{i}\pi-\theta_{12})}(s_1\otimes
\bar{s}_2+\bar{s}_1\otimes s_2), \nonumber
\end{eqnarray}
in agreement with the result first obtained by diagonalization of
the $S^{SG}$-matrix in \cite{KW}. The normalization for two
particles is chosen to be
$N_2^{\Theta}=\frac{\mathrm{i}}{p}M^2\mathcal{C}_p^4$ ($M$ being
the mass of the soliton), in order to have:
$$
f_{SG}^{\Theta}(\theta_1+\mathrm{i}\pi,\theta_1)= 2\pi M^2
(s_1\otimes \bar{s}_2+\bar{s}_1\otimes s_2).
$$
Let us note that the two and four particles form factors were
checked again Feynman graph expansion in \cite{BK}.

\paragraph{\it{Holomorphic components of the stress-energy tensor} $T^{\pm\pm}$.}
It is a chargeless operator with $\pm2$ Lorentz spin. Its
$p$-function reads \cite{BK}:
\begin{eqnarray}
p_{2n}^{T^{\pm,\pm}}=\left(\sum_{i=1}^{n}
e^{\pm\theta_i}\right)\left(\sum_{i=1}^{n}e^{\mp\theta_i}\right)^{-1}
p_{SG}^{\Theta}(\theta,u). \label{TTheta}
\end{eqnarray}
This relation ensures the conservation laws:
$$
\partial_+ \Theta=\partial_-T^{++},\quad \partial_- \Theta=\partial_+ T^{--}.
$$
The two particles form factors read:
\begin{eqnarray}
f_{SG}^{T^{\pm,\pm}}(\theta_{12})= -\frac{2\pi
N_2^{T}}{\mathcal{C}_p^4}f(\theta_{12})e^{\pm(\theta_1+\theta_2)}\frac{\cosh
\frac{\theta_{12}}{2}}{\sinh\frac{1}{2p}(\mathrm{i}\pi-\theta_{12})}(s_1\otimes
\bar{s}_2+\bar{s}_1\otimes s_2), \nonumber
\end{eqnarray}
and
$N_2^{T}=-\frac{\mathrm{i}}{p}M^2\mathcal{C}_p^4=-N_2^{\Theta}$.

\paragraph{\it RSOS restriction \cite{L-RS,RS}.}
The RSOS restriction describes the $\Phi_{1,3}$-perturbations
 of minimal models of CFT \cite{Z} for rational values of $p$.
In particular, we remind that $S_{3}^{RSOS}=-1$ is the Ising
$S$-matrix. Form factors in this model can be
 directly obtained from those of the SG model, as explained in
 \cite{RS}. For the trace operator, the RSOS procedure consists
 of 'taking the half' of the $p$-function\footnote{The rationale behind this is explained in \cite{RS}.}
 (\ref{trace}), such that its $p$-function reads:
\begin{eqnarray}
p_{RSOS}^{\Theta}(\theta,u)=-\sum_{i=1}^{2n}e^{-\theta_i}\sum_{m=1}^{n}e^{u_m},
\label{prsos}
\end{eqnarray}
then we should modify the Bethe ansatz state:
\begin{eqnarray}
\tilde{\Psi}^{p}_{\epsilon_1 \epsilon_2 \dots
\epsilon_{2n}}(\theta,u)\equiv e^{\frac{1}{2p}\sum_i \epsilon_i
\theta_i} \Psi^{p}_{\epsilon_1 \epsilon_2 \dots
\epsilon_{2n}}(\theta,u), \quad \epsilon_i=\pm,\;
\sum_{i=1}^{2n}\epsilon_i=0. \label{ms}
\end{eqnarray}
The two particles form factor of the energy momentum tensor read
explicitly:
\begin{eqnarray}
f_{RSOS}^{\Theta}(\theta_{12})= \frac{2\pi
N_2^{\Theta}}{\mathcal{C}_p^4}f_p(\theta_{12})\frac{\cosh
\frac{\theta_{12}}{2}}{\sinh\frac{1}{p}(\mathrm{i}\pi-\theta_{12})}(e^{\frac{\mathrm{i}\pi}{2p}}s_1\otimes
\bar{s}_2 + e^{-\frac{\mathrm{i}\pi}{2p}}\bar{s}_1\otimes s_2),
\label{thrsos}
\\
f_{RSOS}^{T^{\pm \pm}}(\theta_{12})= -\frac{2\pi
N_2^{\Theta}}{\mathcal{C}_p^4}f_p(\theta_{12})e^{\pm(\theta_1+\theta_2)}\frac{\cosh
\frac{\theta_{12}}{2}}{\sinh\frac{1}{p}(\mathrm{i}\pi-\theta_{12})}(e^{\frac{\mathrm{i}\pi}{2p}}s_1\otimes
\bar{s}_2 + e^{-\frac{\mathrm{i}\pi}{2p}}\bar{s}_1\otimes s_2).
\label{trsos}
\end{eqnarray}
and the normalization for two particles is chosen to be
$N_2^{\Theta}=\frac{2}{p}\mathrm{i}M^2\mathcal{C}_p^4$, in order
to have:
$$
f_{RSOS}^{\Theta}(\theta_1+\mathrm{i}\pi,\theta_1)= 2\pi
M^2(e^{\frac{\mathrm{i}\pi}{2p}}\; s_1\otimes \bar{s}_2 +
e^{-\frac{\mathrm{i}\pi}{2p}}\; \bar{s}_1\otimes s_2).
$$
One can obtain the form factors of the primaries $\Phi_{1,s}$
using the identification
 $\Phi_{1,s}\sim e^{\mathrm{i}\frac{(s-1)}{2}\beta \varphi_{SG}(x)}$,
 (thus using the corresponding $p$-function) then by twisting the Bethe
 ansatz state like in (\ref{ms}) \cite{RS}.

\section{Massless perturbation of the coset models}
\label{two} We consider the massless flows \cite{CSS} from the UV
coset model \cite{GKO} $su(2)_{k+1} \otimes su(2)_k
/su(2)_{2k+1}$, with central charge
$$
 c_{\textrm{\tiny UV}}=
\frac{3k(k+1)(2k+5)}{(k+2)(k+3)(2k+3)},
$$
 to the IR
coset $su(2)_{k} \otimes su(2)_1 /su(2)_{k+1}$. The latter model
is the minimal model $M_{k+2}$ with central charge
$$
c_{\textrm{\tiny IR}}= \frac{k(k+5)}{(k+2)(k+3)}.
$$
The flow is defined in UV by the relevant operator of conformal
dimension $\Delta=\bar{\Delta}=1-2/(2k+3)$; it
arrives in the IR along the irrelevant operator $T\bar{T}$.\\
In the massless case, the dispersion relations read
$(p^0,p^1)=\frac{M}{2}(e^{\theta},e^{\theta})$ for right movers
and $(p^0,p^1)=\frac{M}{2}(e^{-\theta'},-e^{-\theta'})$ for left
movers, where $M$ is some mass-scale in
 the theory, and $\theta,\theta'$ the rapidity variables. Zero momentum
  corresponds to $\theta\to -\infty$ for right movers and $\theta'\to +\infty$ for left movers.\\
The $S$-matrices for the three different scatterings were found in
\cite{Bernard}: the $RR$ and $LL$ $S$-matrices describe the IR CFT
and are thus given by the RSOS restriction of the Sine-Gordon
$S$-matrix. The parameter $p$ is related to $k$ by $p\equiv k+2$.
The $RL$ scattering is given by\footnote{For the particular cases
$p=3$ and $p=+\infty$, the scattering datas were first proposed in
\cite{Z2} and \cite{ZZ} respectively.}:
\begin{eqnarray}
S_{RL}(\theta-\theta')=\frac{1}{S_{LR}(\theta'-\theta)}=\tanh\left(\frac{\theta-\theta'}{2}-\frac{\mathrm{i}\pi}{4}\right)
\, . \nonumber
\end{eqnarray}
The minimal form-factor in the RL channel satisfies the relation:
$$
f_{RL}(\theta-\theta')=f_{RL}(\theta-\theta'+2\mathrm{i}\pi)S_{RL}(\theta-\theta'),
$$
and its explicit expression is given by
\begin{eqnarray}
f_{RL}(\theta-\theta')=\exp\left(
\frac{(\theta-\theta'-\mathrm{i}\pi)}{4}-\int_{0}^{+\infty}\;
 \frac{dt}{t}\frac{1-\cosh t\left(1-\frac{(\theta-\theta')}{\mathrm{i}\pi}\right)}
 {2\sinh t\cosh \frac{t}{2}}\right).
\nonumber
\end{eqnarray}
Its asymptotic behaviour in the infrared is:
$f_{RL}(\theta-\theta')\sim \mathcal{K}\;
e^{\frac{1}{2}(\theta-\theta'-\mathrm{i}\pi)}$, where
\begin{eqnarray}
\mathcal{K}=\exp{-\frac{1}{2}\int_{0}^{\infty}\frac{dt}{t}\left(\frac{1}{\sinh
t \cosh \frac{t}{2}}-\frac{1}{t}\right)}. \nonumber
\end{eqnarray}

\subsection{Form factors of the trace operator $\Theta$}
\subsubsection{Trace operator in the flow from TIM to IM}
The case $p=3$ corresponds to the massless
 flow with spontaneously broken supersymmetry between the tricritical Ising model
 ($c=7/10$) and the Ising model ($c=1/2$). The form factors of $\Theta$ have non
 vanishing matrix elements on an even number of right
and left particles. They satisfy the residue equation in the $RR$
channel at $\theta_1=\theta_{2r}+\mathrm{i}\pi$
($S_{RR}=S_{LL}=-1$):
\begin{eqnarray}
\lefteqn{\mathrm{res}F^{\Theta}_{2r,2l}(\theta_1,\cdots,\theta_{2r};\theta'_1,\cdots,\theta'_{2l})=} \nonumber\\
&&
-2\mathrm{i}\;F^{\Theta}_{2r-2,2l}(\theta_2,\cdots,\theta_{2r-1};\theta'_1,\cdots,\theta'_{2l})
\left(1-\prod_{k=1}^{2l}S_{RL}(\theta_{2r}-\theta'_k)\right).\nonumber
\end{eqnarray}
A similar relation holds in the $LL$ channel. The first non
vanishing matrix element has two right
 and two left particles, and reads \cite{DMS}:
\begin{eqnarray}
F^{\Theta}_{2,2}(\theta_1,\theta_{2};\theta'_1,\theta'_{2})=
\frac{4\pi
M^2}{\mathcal{K}^4}\sinh\frac{\theta_{12}}{2}\sinh\frac{\theta'_{12}}{2}\prod_{i=1,2,\atop
j=1,2} f_{RL}(\theta_i-\theta'_j). \label{thetaIM}
\end{eqnarray}
The form factors of the trace operator were first computed for a
small number of particles in \cite{DMS} in terms of symmetric
polynomials. A formula valid for an arbitrary number of particles
was then presented in \cite{MS}. Subsequently in \cite{P} it was
noticed that it is possible to simplify
 the formula of \cite{MS} such that the form factors could be rewritten as:
\begin{eqnarray}
\lefteqn{F_{2r,2l}^{\Theta}(\theta_1,\ldots,\theta_{2r};\theta'_1,\ldots,\theta'_{2l})=\frac{4\pi
M^2}{\mathcal{K}^4}}
\nonumber \\
&&
 \times \;\prod_{1\leq i<j\leq 2r}\sinh\frac{\theta_{ij}}{2}
\prod_{1\leq i<j\leq 2l}\sinh\frac{\theta'_{ij}}{2}
\prod_{i,j}f_{RL}(\theta_i -\theta'_j)
Q_{2r,2l}^{\Theta}(\theta_1,\ldots,\theta_{2r};\theta'_1,\ldots,\theta'_{2l}).
\nonumber
\end{eqnarray}
Explicitly:
\begin{eqnarray}
\lefteqn{Q_{2r,2l}^{\Theta}(\theta_1,\ldots,
\theta_{2r};\theta'_1,\ldots,\theta'_{2l})=}\nonumber \\
&& \pi^{r+l-2}\sum_{T \subset S, \atop \#T=r-1}\sum_{T' \subset
S',\atop \#T'=l-1}\prod_{i \in T, \atop k\in
\bar{T}}\phi(\theta_{ik})\prod_{i \in T', \atop  k\in
\bar{T}'}\phi(\theta'_{ik}) \prod_{i \in T, \atop k\in
\bar{T}'}\Phi(\theta_i-\theta'_k) \prod_{i \in T', \atop k\in
\bar{T}}\tilde{\Phi}(\theta_k-\theta'_i),\nonumber
\end{eqnarray}
and
\begin{eqnarray}
\phi(\theta_{ij}) \equiv
\frac{-1}{f_{RR}(\theta_{ij})f_{RR}(\theta_{ij}+\mathrm{i}\pi)}
=\frac{2\mathrm{i}}{\sinh \theta_{ij}}\; ,\quad \phi(\theta'_{ij})
\equiv
\frac{-1}{f_{LL}(\theta'_{ij})f_{LL}(\theta'_{ij}+\mathrm{i}\pi)}
= \frac{2\mathrm{i}}{\sinh \theta'_{ij}}\; ,\label{phi}
\end{eqnarray}
as well as:
\begin{eqnarray}
\Phi(\theta-\theta')\equiv
\frac{S_{RL}(\theta-\theta')}{f_{RL}(\theta-\theta')f_{RL}(\theta-\theta'+\mathrm{i}\pi)}=
 \mathcal{K}^{-2}(1-\mathrm{i}e^{\theta'-\theta}),\quad
\tilde{\Phi}(\theta-\theta') \equiv
\Phi(\theta-\theta'+\mathrm{i}\pi). \label{grandphi}
\end{eqnarray}
Let us note that the normalization of the form factors of the
trace operator in (\ref{thetaIM})
 was chosen in order to match the effective Lagrangian \cite{DMS}:
$$
F^{\Theta}_{2,2}\to -\frac{1}{\pi M^2}f^{T}_{2}f^{\bar{T}}_2,
$$ where $f^{T}_{2}$
 and $f^{\bar{T}}_2$ are the two particles form-factors of the
 right and left components
  of the energy momentum tensor in the thermal Ising model:
\begin{eqnarray}
f^{T}_{2}(\theta_1,\theta_2)=2\mathrm{i}\pi
M^2e^{\theta_1+\theta_2}\sinh\frac{\theta_{12}}{2}, \quad
f^{\bar{T}}_2(\theta'_1,\theta'_2)=2\mathrm{i}\pi
M^2e^{-\theta'_1-\theta'_2}\sinh\frac{\theta'_{12}}{2}.
\end{eqnarray}
This normalization for the trace operator ensured a proper
numerical check of the $c$-theorem in \cite{DMS}.

\subsubsection{Generalizations}
The form factors of the trace operator for $p$ arbitrary were
first obtained in \cite{MS} by M\'ejean and Smirnov in a different
representation from the one presented below. The problem of
normalization
of form factors is not discussed in \cite{MS}, and the numerical checks are not performed.\\
The form factors of $\Theta$ have non vanishing matrix elements on
an even number of right and left particles. They satisfy the
residue equation in the $RR$ channel at
$\theta_1=\theta_{2r}+\mathrm{i}\pi$:
\begin{eqnarray}
\lefteqn{\mathrm{res}F^{\Theta}_{2r,2l}(\theta_1,\cdots,\theta_{2r};\theta'_1,\cdots,\theta'_{2l})=} \nonumber\\
&&
-2\mathrm{i}\;F^{\Theta}_{2r-2,2l}(\theta_2,\cdots,\theta_{2r-1};\theta'_1,\cdots,\theta'_{2l})
\left(1-\prod_{i=2}^{2r-1}S^{RSOS}_p(\theta_{i}-\theta_{2r})
\prod_{k=1}^{2l}S_{RL}(\theta_{2r}-\theta'_k)\right)
e^{R},\nonumber
\end{eqnarray}
where $e^{R}=e^{\frac{\mathrm{i}\pi}{2p}}s_1\otimes
\bar{s}_{2r}+e^{-\frac{\mathrm{i}\pi}{2p}}\bar{s}_1\otimes
s_{2r}$.
 A similar relation holds in the $LL$ channel. \\
 The lowest form factor for the trace operator contains two right
  and two left particles; it reads:
\begin{eqnarray}
\lefteqn{F^{\Theta}_{2,2}(\theta_1,\theta_2;\theta'_1,\theta'_2) }\nonumber\\
&& = \frac{16\pi M^2 }{p^2\mathcal{K}^4}\;
f_p(\theta_{12})f_p(\theta'_{12})\prod_{i=1,2 \atop
j=1,2}f_{RL}(\theta_{i}-\theta'_{j})
\frac{\cosh\frac{\theta_{12}}{2}\cosh\frac{\theta'_{12}}{2}\;
e^{R}\otimes
e^L}{\sinh\frac{1}{p}(\theta_1-\theta_2-\mathrm{i}\pi)\sinh\frac{1}{p}(\theta'_1-\theta'_2-\mathrm{i}\pi)}\nonumber \\
&& =-\frac{1}{\pi \mathcal{K}^4 M^2}\prod_{i=1,2 \atop
j=1,2}f_{RL}(\theta_{i}-\theta'_{j})\;e^{\theta'_1+\theta'_2-\theta_1-\theta_2}\;
f^{T^{++}}_{RSOS}(\theta_{12})f^{T^{--}}_{RSOS}(\theta'_{12}),
\label{traceRSOS}
\end{eqnarray}
where $f^{T^{\pm,\pm}}_{RSOS}(\theta_{12})$ is defined in equation
(\ref{trsos}).\\
 We make the following ansatz for the solution of
the form factors equations with the first
 recursion step given by (\ref{traceRSOS}):
\begin{eqnarray}
\lefteqn{F^{\Theta}_{2r,2l}(\theta_1,\dots,\theta_{2r};\theta'_{1},\dots,\theta'_{2l})=
\frac{N^{\Theta}_{2r}N^{\Theta}_{2l}}{M^2\pi\mathcal{K}^4}\;\prod_{1\le
i<j\le 2r}f_p(\theta_{ij})\prod_{1\le i<j\le 2l}f_p(\theta'_{ij})
\prod_{i,j}f_{RL}(\theta_i-\theta'_{j})
} \nonumber \\
&& \times \int_{C_{\theta}}\prod_{m=1}^r du_m\;
h_{RR}(\theta,u)p_{RSOS}^{T^{++}}(\theta,u)
 \tilde{\Psi}^{p}(\theta,{u})\; \int_{C_{\theta'}}\prod_{m=1}^l dv_m\; h_{LL}(\theta',v)
p_{RSOS}^{T^{--}}(\theta',v) \tilde{\Psi}^{p}(\theta',{v})\nonumber \\
&& \times \; \mathcal{M}_{2r,2l}(\theta,\theta',{u},{v}),\nonumber
\end{eqnarray}
The $p$-functions $p_{RSOS}^{T^{\pm,\pm}}(\theta,u)$ are defined
through equation (\ref{prsos}) and the analogue of equation
(\ref{TTheta}). The Bethe ansatz state $\tilde{\Psi}^{p}(\theta,u)$ was introduced in (\ref{ms}).\\
At $\theta_{1}=\theta_{2r}+\mathrm{i}\pi$, each of the $r$
integration contours gets
 pinched at $\theta_{2r},\theta_{2r}\pm \mathrm{i}\pi$.
For this reason we introduced the function
$\mathcal{M}_{2r,2l}(\theta,\theta',u,v)$ that
 should satisfy the properties at $\theta_{1}=\theta_{2r}+\mathrm{i}\pi$:
\begin{itemize}
\item
 $u_r=\theta_{2r}$:
\begin{eqnarray}
\lefteqn{\mathcal{M}_{2r,2l}(\theta_1,\dots,\theta_{2r};\theta';u_1,\dots
,u_r;v)=}
\nonumber \\
&& \mathcal{M}_{2r-2,2l} (\theta_2,\dots,\theta_{2r-1};\theta';
u_1,\dots, u_{r-1};v)
 \prod_{k=1}^{2l}\Phi(\theta_{2r}-\theta'_k).
\nonumber
\end{eqnarray}
\item
 $u_r=\theta_{2r}\pm \mathrm{i}\pi$:
\begin{eqnarray}
\lefteqn{\mathcal{M}_{2r,2l}(\theta_1 \dots \theta_{2r};\theta';
u_1 \dots u_r;v)=}\nonumber \\
&& \mathcal{M}_{2r-2,2l} (\theta_2,\dots,\theta_{2r-1};\theta';
u_1,\dots, u_{r-1};v)
\prod_{k=1}^{2l}\tilde{\Phi}(\theta_{2r}-\theta'_k).\nonumber
\end{eqnarray}
\end{itemize}
Similar relations hold in the $LL$ channel.\\
We introduce
 the sets $S=(1,\dots,2r)$ and $S'=(1,\dots,2l)$, as
well as the subsets $T\subset S\;,T'\subset S'$ and $\bar{T}\equiv
S\backslash T,\; \bar{T'}\equiv S'\backslash T'$. These subsets
have the number of elements: $\#T=r-1$, $\#\bar{T}=r+1$
\begin{eqnarray}
T = \{i_1<i_2<\dots<i_{r-1}\},\quad \bar{T} =
\{k_1<k_2<\dots<k_{r+1}\}, \nonumber
\end{eqnarray}
and $T',\bar{T}'$ are defined similarly. We propose the following
function:
\begin{eqnarray}
\mathcal{M}_{2r,2l}(\theta,\theta',{u},{v})=\frac{4\;
\mathrm{i}^{r+l-2}}{\sum_{i=1}^{2r}
 e^{\theta_i}\sum_{i=1}^{2l} e^{-\theta'_i}} \sum_{T \subset S, \atop \#T=r-1}\sum_{T' \subset S',\atop
\#T'=l-1} \frac{\prod_{k,l\in \bar{T}\atop k<l
}\cos\frac{\theta_{kl}}{2\mathrm{i}}} {\prod_{i\in T,\atop k\in
\bar{T}}\sin\frac{\theta_{ki}}{2\mathrm{i}}}e^{\frac{1}{2}\sum\theta_{ik}}
\frac{\prod_{k,l\in \bar{T'}\atop
k<l}\cos\frac{\theta'_{kl}}{2\mathrm{i}}}{\prod_{i\in T',\atop
k\in \bar{T}'}
\sin\frac{\theta'_{ki}}{2\mathrm{i}}}e^{\frac{1}{2}\sum\theta'_{ki}}\nonumber
\\\times \;
 \frac{\prod_{i\in T,\atop m=1,\dots
,r}\cos\frac{\theta_i-u_m}{2\mathrm{i}} \prod_{i\in T',\atop
m=1,\dots ,l}\cos\frac{\theta'_i-v_m}{2\mathrm{i}}}{\prod_{1\le
m<n\le r} \cos \frac{u_m-u_n}{2\mathrm{i}}\prod_{1\le m<n\le
l}\cos \frac{v_m-v_n}{2\mathrm{i}}}\prod_{i \in T, \atop k\in
      \bar{T}'}\Phi(\theta_i-\theta'_k) \prod_{i \in T', \atop k\in
      \bar{T}}\tilde{\Phi}(\theta_k-\theta'_i)\;
\nonumber . \label{M}
\end{eqnarray}
In particular, for two right movers and two left movers:
$\mathcal{M}_{2,2}(\theta_1,\theta_2;\theta'_1,\theta'_2;u;v)=e^{-\theta_1-\theta_2+\theta'_1+\theta'_2}$.\\
A similar function was first obtained in \cite{MS} in a different
representation\footnote{The integral representation for the form
factors proposed in \cite{MS} is different from ours, as it
contains two integration variables less (one less for each
right/left channel). We did not manage to prove that the
representation for our form factors coincides with the one of
\cite{MS}.}, and was used in a very different context in
\cite{BKS}. We present it here in what we believe to be a simpler
expression.
 It should also be compared with a very resemblant function obtained in \cite{P2} in the massive case.\\
Finally, the following recursion relations hold \cite{BFKZ}:
$$
N_{2r}^{\Theta}=N_{2r-2}^{\Theta}(f_p^{min}(0))^2/4\pi r, \quad
N_{2l}^{\Theta}=N_{2l-2}^{\Theta}(f_p^{min}(0))^2/4\pi l.
$$
We set the normalization of the $p$-function
$N_{2}^{\Theta}=\frac{2\mathrm{i}M^2\mathcal{C}_p^4}{p}$ in order
to have in the infrared the same relation as in the flow from TIM
to IM\footnote{We are grateful to G.~Delfino for explaining to us
how to normalize properly the form factors of the trace
operator.}:
\begin{eqnarray}
F^{\Theta}_{2,2}\to -\frac{1}{\pi M^2}f^{T}_{2}f^{\bar{T}}_2,
\label{irlimit}
\end{eqnarray}
where $f^{T}_{2}$
 and $f^{\bar{T}}_2$ are the two particles form-factors of the
 right and left components
  of the energy momentum tensor in the $M_{k+2}$ minimal model
 perturbed by $\Phi_{1,3}$ -see (\ref{trsos})-,
   themselves normalized such that:
$$f^{T^{++}}(\theta_1+\mathrm{i}\pi,\theta_1)
=2\pi M^2e^{2\theta_1}e^R\; , \quad
f^{T^{--}}(\theta'_1+\mathrm{i}\pi,\theta'_1)=2\pi
M^2e^{-2\theta'_1}e^L.
$$
In \cite{MS}, it is shown that for an arbitrary number of
particles the form factors of the trace operator reproduce the
direction of the flow: $F^{\Theta}_{2r,2l}\to -\frac{1}{\pi
M^2}f^{T}_{2r}f^{\bar{T}}_{2l}$. We verified with Mathematica that
this is true on our representation, but this check was done for a
small number of particles only.

\section{Numerical results}
\label{three}
The knowledge of the form factors
of the trace of the stress energy tensor allows to estimate the
variation of the central charge along the flow by means of the
"$c$-theorem" sum rule \cite{Zamolodchikov:1986gt,Cardy:1988tj}:
\begin{eqnarray}
\Delta c= c_{\textrm{\tiny UV}}-c_{\textrm{\tiny
IR}}=\frac{3}{2}\int_{0}^{\infty}dr\; r^3 \langle
\Theta(r)\Theta(0)\rangle.\label{cth}
\end{eqnarray}
Since in the massless case any correlation function can be
represented by its spectral expansion
\begin{eqnarray}
\lefteqn{\langle O(x)O(0) \rangle =} \label{int} \\
&& \sum_{r,l=0}^{\infty}\frac{1}{r!l!}\int_{-\infty}^{+\infty}
\frac{d\theta_1\dots d\theta_r\;d\theta'_1\dots
d\theta'_l}{(2\pi)^{r+l}}
|F_{r,l}^{O}(\theta_1,\ldots,\theta_{r};\theta'_1,\ldots,\theta'_{l})|^2
e^{-\frac{Mr}{2}\left(\sum_{j=1}^{r}e^{\theta_j}+\sum_{j=1}^{l}e^{-\theta'_j}
\right)}, \nonumber
\end{eqnarray}
the computation of $\Delta c $ turns out to be a non trivial check
for the form factors $F^{\Theta}_{2r,2l}$ (at least for the first
few of them).

In the past years, the calculation of $\Delta c$ using form
factors has been done in various massive theories, providing
accurate results even within the two-particle approximation.
However in \cite{CAF}, and more recently in \cite{P2} within the
construction of form factors for the (massive) $SS$ model and its
RSOS restrictions, significantly large discrepancies have been
observed comparing the computation of the central charge by means
of the $c$-theorem and the corresponding exact results. In
\cite{P2}, within the two-particle approximation and for a given
subset of the parameters of the model, the deviations were of
about 20-25\%. On the other hand, the massless flow from the
Tricritical Ising to the Ising model provided up to now the unique
massless case known where numerical checks for the $c$-theorem
were performed: the numerical results obtained in \cite{DMS} by
Delfino, Mussardo and Simonetti are unexpectedly accurate, as the
leading contribution (four particles) was enough to obtain the
98\% of the exact result. This is a very remarkable situation, as
there is absolutely no reason to expect in the massless case that
the leading contribution gives any good approximation at all to
the correlation function: at a given energy, one cannot say what
number of particle processes contribute.

In the present case, we deal with a one parameter family of
massless flows, whose variation of the central charge is given by
\begin{eqnarray}
\Delta c_k^{\textrm{\tiny exact}}= c_{\textrm{\tiny
UV}}-c_{\textrm{\tiny IR}} = \frac{4 k^2}{2 k^2 +9k +9},
\label{dck}
\end{eqnarray}
(we recall that $k=1$ corresponds to the flow from TIM to IM).
Moreover, it is of interest to compare the accuracy of the
numerical results for the variation of the central charge in the
massless case against the accuracy obtained in the associated
massive coset model \cite{P2}, where the UV CFT is the same as the
one considered in the previous massless flows, and whose central
charge is given by
\begin{eqnarray}
c_k^{\textrm{\tiny exact}}= c_{\textrm{\tiny UV}} = \frac{3
k(k+1)(2k+5)}{(k+2)(k+3)(2k+3)}. \label{ck}
\end{eqnarray}
Let us start with the massless flow: in the following we will
consider the correlation function truncated to 4 particles, with
the use of the form factor (\ref{traceRSOS}); higher form factors
are very difficult to compute numerically because the scattering
is non-diagonal for $k>1$ (the 6-particles contribution for the
diagonal case $k=1$, corresponding to the TIM $\to$ IM flow, was
considered in \cite{DMS}, starting with the formula
(\ref{thetaIM}) for four particles).
\begin{center}
\begin{table}
\begin{center}
\begin{tabular}{|c|l|l|r|}
\hline
$k$ & $\Delta c_k^{(4)}$ & $\Delta c_k^{\textrm{\tiny exact}}$ & \% dev.\\
\hline \hline
$-1$ & $2.061(1)$ & $2$ & $3 \% $\\
\hline \hline
$1$ & $0.197(2)$ & $0.2$ & $2 \% $\\
$2$ & $0.433(2)$ & $0.457143 \dots$ & $5 \% $\\
$3$ & $0.608(2)$ & $0.666666 \dots$ & $9 \% $\\
$4$ & $0.729(3)$ & $0.831169 \dots$ &$12 \% $ \\
$6$ & $0.881(3)$ & $1.066666 \dots$ & $17 \% $\\
$8$ & $0.967(3)$ & $1.224880 \dots$ &$21 \% $ \\
$10$ & $1.012(4)$ & $1.337792  \dots$ & $24 \% $\\
$20$ & $1.125(4)$ & $1.617795  \dots$ & $30 \% $\\
$50$ & $1.139(5)$ & $1.831837  \dots$ & $38 \% $\\
\hline \hline
$\infty$ &  $1.130(5)$  & $2$ &  $43 \%$ \\
\hline
\end{tabular}
\end{center}
\caption{Massless flow~-~four particles approximation (the number
in bracket in the second column is the statistical error on the
previous digit due to the Montecarlo integration).} \label{tab1}
\end{table}
\end{center}
Within such an approximation we obtained the numerical estimates
collected in Table \ref{tab1} where the comparison with the
expected results (\ref{dck}) is also given. In order to explain
the observed deviations from the exact values of $\Delta
c_k^{\textrm{\tiny exact}}$, it is worth remembering that the
conformal dimension of the perturbing operator (in the UV) is
given by
\begin{eqnarray}
\Delta = \bar{\Delta}=1-\frac{2}{2 k + 3} \nonumber
\end{eqnarray}
which becomes marginally relevant in the limit $k \to \infty$.
Hence, in such
a limit, the UV behaviour of the correlator in the sum rule tends
to $r^{-4}$ (also logarithmic corrections may appear), making the
convergence of the integral (\ref{cth}) weaker as $k$ is increased.

As a consequence, at large but finite $k$ the suppression in the
UV region ({\it i.e.} where the form factors approximation is
worse) becomes very weak giving rise to the observed deviations.
In principle, such an argument can also explain the reason why the
four-particle approximation shows large discrepancies at
relatively small values of $k$.

In order to illustrate this point let us consider the following
example: when $k=4$ we have $\Delta = 0.818 \dots$ and a
corresponding deviation of the 12\% which can be compared with the
case $k=1$, where the deviation is of about 2\% and one has
$\Delta = 0.6$. Hence it is likely to conjecture that the previous
mechanism is a good candidate to be the responsible for the poor
quality of the numerical estimates of $\Delta c_k$. Such an
interpretation also implies that it is of a moderate interest to
work out the contributions due to 6-particles form factors: we
expect that they will add a quantitative correction to our results
without changing the global picture outlined above.

A similar behaviour has been observed \cite{P2} in the massive
flow originating from the coset model\footnote{This coset model is
obtained by RSOS restriction of the $SS$ model \cite{Fateev}.}
$su(2)_{k+1} \otimes su(2)_{k} / su(2)_{2k+1}$ perturbed by the
same operator as before with conformal dimension $\Delta$ (the
sign of the perturbation is different). In the massive case, the
lowest form factor needed for numerical tests contains 2
particles. As one can see in Table \ref{tab2}, the numerical
determination of the central charge $c_k^{(2)}$ given by
({\ref{ck}) becomes worse as $k$ is increased (we observe that in
this case the discrepancy is not as large as in the massless
flow\footnote{Such a difference is due to the presence of the
exponential factor $\exp(-m R \cosh \theta)$ inside the spectral
expansion for correlators
 in massive theories, see
\cite{DMS}.}). Since the perturbing operator is the same, we
expect that the reason of such a decreasing of the precision is
due to the same mechanism as before.

As a side remark it is interesting to notice that the limit $k \to
\infty$ seems to resemble the case of the $SU(2)$ Thirring model
(which is nothing but the Sine-Gordon model at the marginal point
$\beta^2=8\pi$): even if the perturbation is marginally relevant,
the two-particle approximation to the sum rule gives a finite
result. The very important difference is that in the latter case
the numerical computation \cite{Delfino,P2} gives $c^{(2)}=0.987$
which is unexpectedly near to the exact one $c^{\textrm{\tiny
exact}}=1$ (instead of being very far as in the case of coset
models, where we have $ 43 \%$ and $21 \%$ for the massless and
massive models respectively). The previous examples shows that one
has to be very cautious in interpretating the results which come
from the form factor approximation to the $c$-theorem sum rule in
the case of marginal perturbations (we refer the reader to
\cite{P2} for additional numerical examples in the
$SU(2)_k$-Thirring model)\footnote{More generally, even within the
(massive) minimal models $M_{k+2}$ one observes a loss of
precision when increasing the parameter $k$ \cite{P2}; however it
is nothing comparable with the phenomenon observed here.}.

We find it interesting to present the results for the case $k=-1$,
which corresponds this time to a {\it non-unitary} flow from a UV
fixed point with $c_{\textrm{\tiny UV}}=0$ to a IR CFT with
$c_{\textrm{\tiny IR}}=-2$; the result in Table \ref{tab1} shows
that the accuracy of the estimate for the central charge is good,
and the fact that the approximation to the exact value comes from
above is in agreement with the non-unitarity of the flow.
\begin{center}
\begin{table}
\begin{center}
\begin{tabular}{|c|l|l|r|}
\hline
$k$ & $ c_k^{\textrm{(2)}}$ & $ c_k^{\textrm{\tiny exact}}$ & \% dev.\\
\hline \hline
$1$ & $0.6988$ & $0.7$ & $0.17 \% $\\
$2$ & $1.1429$ & $1.157142 \dots$ & $1.2 \% $\\
$3$ & $1.4306$ & $1.466666 \dots$ & $2.4 \% $\\
$4$ & $1.6268$ & $1.688311 \dots$ &$3.6 \% $ \\
$6$ & $1.8702$ & $1.983333 \dots$ & $5.7 \% $\\
$8$ & $2.0105$ & $2.170334\dots$ &$7.4 \% $ \\
$10$ & $2.0994$ & $2.299331  \dots$ & $8.7 \% $\\
$20$ & $2.2764$ & $2.605938  \dots$ & $12.6 \% $\\
$50$ & $2.3614$ & $2.829660  \dots$ & $17.4 \% $\\
\hline \hline
$\infty$ &  $2.3863$  & $3$ &  $21 \%$ \\
\hline
\end{tabular}
\end{center}
\caption{Massive flow~-~two particles approximation} \label{tab2}
\end{table}
\end{center}
Finally, we used the 4-particle form factors to compute in the
massless case the correlation function $\langle \Theta(x)
\Theta(0) \rangle$. In the figures $1$-$4$ we compared such an
approximation of the correlator with its power-law behaviour at
both the UV and IR fixed points for different values of $k$
($k=1$, $2$, $3$, $10$; the case $k=1$ was already considered in
\cite{DMS}). One may observe that the relation (\ref{irlimit})
plugged into the integral (\ref{int}) followed by a simple change
of variables insures the IR behaviour of the correlation function
taking into account the four particles contribution only,
\begin{eqnarray}
\displaystyle
 \langle \Theta(x) \Theta(0) \rangle \mathop{\sim}_{r\to
\infty}\left(\cos \frac{2\pi}{k+2}+1\right)\times
\frac{2J_{k+2}}{\pi^2 M^4r^{8}}, \nonumber
\end{eqnarray}
where
\begin{eqnarray}
J_{k+2}= \left( \frac{3 I_{k+2}}{(k+2)^2} \right)^2, \ \ \ \ \
I_{k+2}= 4\int_{-\infty}^{+\infty} d \theta  \left| \frac{ f_{k+2}
(\theta)} {\sinh \frac{1}{k+2} (\mathrm{i} \pi - \theta) \cosh
\frac{\theta}{2} } \right|^2.\nonumber
\end{eqnarray}
As for the UV behaviour, it is given by the following conformal
OPE:
\begin{eqnarray}
\langle \Theta(x) \Theta(0) \rangle \mathop{\sim}_{r\to 0} M^4
\left( \frac{8 \pi}{2k+3} \alpha_k  \right)^2  \; (M r)^{-4
\frac{2k+1}{2k+3}}, \nonumber
\end{eqnarray}
where the constants $\alpha_k$ can be extracted from \cite{fateev}
(Eq. (2.2) with $h=n=2$ and $l=k+1$)
\begin{eqnarray}
\alpha_k = \frac{1}{\pi} \frac{k (k+1)}{2k+1} \left[ \frac {\Gamma
\left( \frac{1}{2k+3}  \right) \Gamma \left( \frac{3}{2k+3}
\right)
 }{\Gamma \left( \frac{2k+2}{2k+3} \right) \Gamma \left( \frac{2k}{2k+3} \right)}
 \right]^{-\frac12} \times
\left[\frac{\pi}{4} \frac {\Gamma \left( \frac{2k+3}{2} \right)}
{\Gamma \left( \frac{k+2}{2}\right) \Gamma \left(
\frac{k+3}{2}\right)} \right]^{\frac{4}{2k+3}}. \nonumber
\end{eqnarray}

Looking at the diagrams $1$-$4$ one can clearly notice the
difference between the expected UV power-law behaviour and the
slope given by the truncated form factor expansion. As a
consequence, the combined effect of such a difference in the UV
behaviour and the above mentioned weak suppression in the integral
(\ref{cth}) can be considered as the cause of the discrepancy
between $c$-theorem and exact results.

Remarkably enough, it appears on the charts that at the condition
of not going too far in the UV, the four particles truncation
seems to give a reasonably reliable approximation to the
correlation function all the way from the intermediate region to
the (exact) IR region.

\section*{Concluding remarks}
In this work we constructed form factors of the trace operator in
the massless flow between the coset model $su(2)_{k+1} \otimes
su(2)_k /su(2)_{2k+1}$ and the $ M_{k+2} $ minimal model. We
mimicked the construction of form factors of the trace operator in
the massless flow TIM $\to$ IM obtained in \cite{DMS}: in the
section 2.1.2, we looked for a solution of the residue equation
obtained by replacing the $RR$ and $LL$ $S$-matrices
$S^{RSOS}_3=-1$ by the $S$-matrix $S^{RSOS}_{k+2}$, that should
match the desired properties in the IR. The form factor of the
trace of the stress-energy tensor $\Theta$ with four particles was
used to compute both the (truncated) correlation function $\langle
\Theta(x) \Theta(0) \rangle$ and the variation of the central
charge (by means of the ``$c$-theorem'' sum rule). Since for the
latter we observed a large discrepancy with respect to the exact
results, we carefully analyzed the problem, giving a plausible
explanation for the phenomenon. Nevertheless, it appears on the
diagrams 1-4 that the truncation to 4 particles for $\langle
\Theta(x) \Theta(0) \rangle$ still gives not too bad an
approximation to the full correlation function, at the condition
that one does not go
too far in the UV.\\
In our next article \cite{GP2}, we will generalize the
construction of form factors of the magnetization operator and the
energy operator in the massless flow TIM $\to$ IM \cite{DMS} to
the whole family of flows.

\section*{Acknowledgments}
We thank G.~Delfino for useful discussions. Both authors were
supported by the
 Euclid Network HPRN-CT-2002-00325. The work of P.G. was
also supported by the COFIN ``Teoria dei Campi, Meccanica
Statistica e Sistemi Elettronici'', and B.P. was supported by a
Linkage International Fellowship of the Australian Research
Council.

\newpage

%\begin{figure}
%\centerline{\psfig{figure=k_meno1.eps,width=0.7\textwidth}}
%\caption{ Flow $k=-1$}
%\label{k meno1}
%\end{figure}
\begin{figure}
\centerline{\psfig{figure=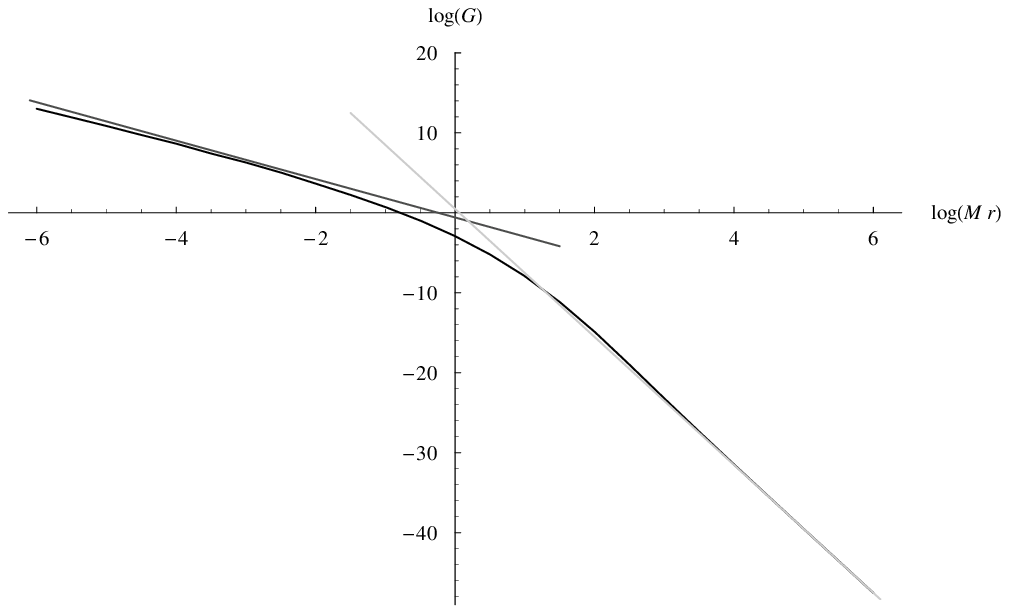,width=0.7\textwidth}}
\caption{Flow $k=1$, logarithmic plot of the correlator
$G(r)=\langle \Theta(x) \Theta(0) \rangle$ (black line) together
with both the IR (grey line) and the UV (dark-grey line)
behaviours.} \label{k_1}
\end{figure}
\begin{figure}
\centerline{\psfig{figure=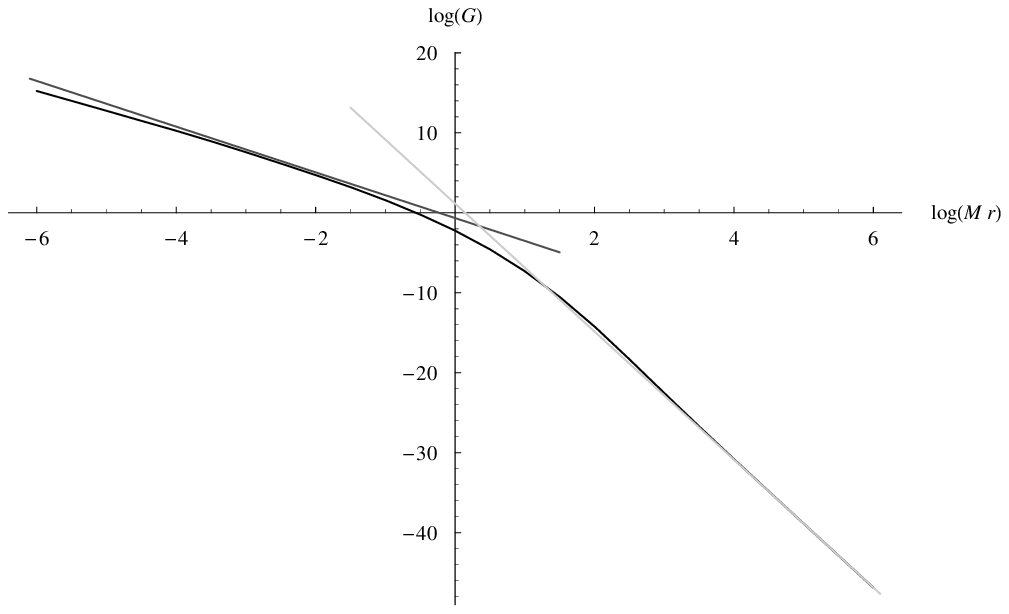,width=0.7\textwidth}}
\caption{Flow $k=2$, logarithmic plot of the correlator
$G(r)=\langle \Theta(x) \Theta(0) \rangle$ (black line) together
with both the IR (grey line) and the UV (dark-grey line)
behaviours.} \label{k_2}
\end{figure}
\begin{figure}
\centerline{\psfig{figure=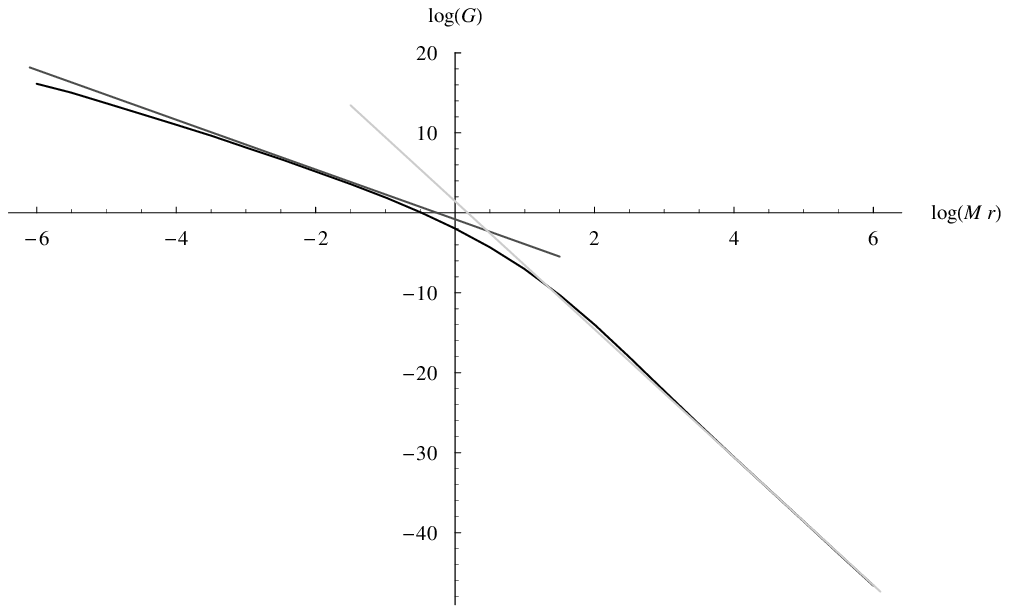,width=0.7\textwidth}}
\caption{Flow $k=3$, logarithmic plot of the correlator
$G(r)=\langle \Theta(x) \Theta(0) \rangle$ (black line) together
with both the IR (grey line) and the UV (dark-grey line)
behaviours.} \label{k_3}
\end{figure}
\begin{figure}
\centerline{\psfig{figure=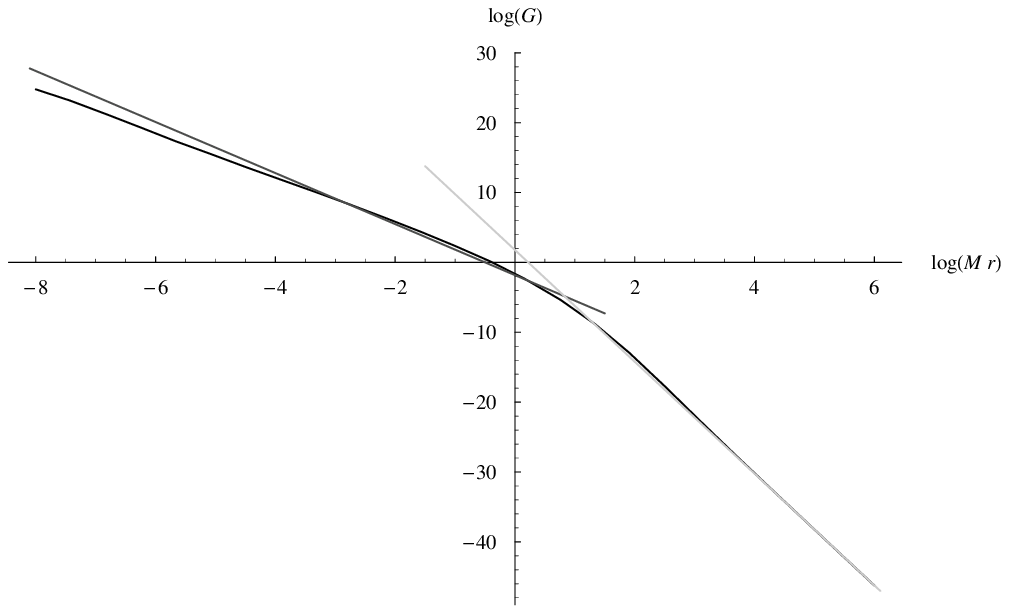,width=0.7\textwidth}}
\caption{Flow $k=10$, logarithmic plot of the correlator
$G(r)=\langle \Theta(x) \Theta(0) \rangle$ (black line) together
with both the IR (grey line) and the UV (dark-grey line)
behaviours.} \label{k_10}
\end{figure}

\end{document}